# Bridging physics-based and equivalent circuit models for lithium-ion batteries

Zeyang Geng[a], Siyang Wang[b], Matthew J. Lacey[c], Daniel Brandell[d], Torbjörn Thiringer[a]

[a]Chalmers University of Technology, Göteborg, Sweden

[b]Mälardalen University, Västerås, Sweden

[c]Scania, Södertälje, Sweden

[d]Uppsala University, Uppsala, Sweden

**Abstract**

In this article, a novel implementation of a widely used pseudo-two-dimensional (P2D) model for lithium-ion battery simulation is presented with a transmission line circuit structure. This implementation represents an interplay between physical and equivalent circuit models. The discharge processes of an NMC-graphite lithium-ion battery under different currents are simulated, and it is seen the results from the circuit model agree well with the results obtained from a physical simulation carried out in COMSOL Multiphysics, including both terminal voltage and concentration distributions. Finally we demonstrated how the circuit model can contribute to the understanding of the cell electrochemistry, exemplified by an analysis of the overpotential contributions by various processes.

**Keywords:** Lithium-ion battery, Pseudo-two-dimensional model, Transmission line model, Overpotential

## 1. Introduction

For lithium-ion batteries, mathematical models not only constitute tools to estimate the performance of different battery components, as well as the cell or the battery pack, but also provide tools to strengthen the understanding of many physical properties, which determine the electrochemical response during the battery operation. An adequate model can be used to both interpret experiment results [1] and offer estimations of quantities that cannot be easily accessed through measurement, for example local overpotentials, capacity losses, morphology changes [2], internal temperature fluctuations and the growth of the solid electrolyte interphase (SEI) layer [3, 4, 5]. Modelling can aid battery diagnostics, and thereby help to prevent premature ageing of the cells [6, 7, 8].

From atomistic scale to system scale, there exist battery models with different complexities depending on the application. In battery management systems, empirical circuit models are commonly used which can fairly accurately describe the dynamic behavior of batteries and can be computed in real time with on-board microprocessors. However, the quality of a circuit model is directly dependent on parameterization data which requires massive amount of experiments at different state of charges (SOC) and temperatures [9]. Moreover, the effectiveness of the model decreases with the battery being aged since all the parameters are affected by the state of health (SOH). It is also a challenge to transfer these models to different generations of cells or types of batteries, which make them reliable for only a limited segment of battery devices. In many cases, Kalman filters are used together with the empirical circuit models to improve the accuracy [10, 11]. While empirical circuit models can illustrate the dynamic behavior of batteries with a transparent structure, such a structure gives very little or no information about the physical parameters actually governing the behavior of the system.

A physics-based approach can instead be employed using the first principles-based lithium-ion battery model that was developed by Newman, Doyle and Fuller [12, 13] and has been implemented into a number of commercial softwares, e.g. COMSOL Multiphysics. Newman's model is a Pseudo-two-Dimensional (P2D) model consisting of a set of partial differential equations (PDEs). The PDEs are typically solved with the finite element method [14] or the finite volume method [15] which leads to a high computational load. As a consequence of the long computational time, there has been a great effort on simplifying the P2D model into single particle models [16, 17] or reduced-order P2D models with polynomial approximations [18]. These simplifications increase the calculation efficiency but also limits the conditions where the models are valid [19]. Furthermore, the physics-based models with complex mathematical equations remains challenging to use and interpret for researchers without a solid mathematics or physics background.

To reduce the gap between the empirical equivalent circuit models and the physics-based models, different physics-based circuit models have recently been implemented in several different ways. One simple approach is to use the transmission line structure with an elementary resistor network but without any link to the mass transport process in the cell [20, 21]. One further step, employed by Sato et al., was to connect the elements in the circuit model with physical principles [22], but then the current distribution within the electrode is ignored in their model. A recent interesting approach used by Li et al. [23] is to construct a physics-based equivalent circuit model with passive electrical components, which is fast and accurate, however, the number of circuit elements, voltage sources and transformers make the model somewhat less easy to comprehend and implement.

So far, the physics-based circuit models reported in literature are either simplified [20, 21, 22] and thus cannot fully capture the battery cell behaviors described in Newman's P2D model, or with an advanced structure [23], somewhat too much exceeding the simplicity of the usual transmission line models. These facts are crucial flaws, and thus forms an important research gap, i.e. to have an easy understandable, but still accurate, circuit-based model of a Li-Ion cell.

To bridge this gap, in this work we present a novel implementation for Newman's P2D model with a transmission line circuit structure. The circuit structure not only solves the current distribution with the mesh current method, but also offers a clear visual illustration of the P2D model. In this implementation, the physics-based electrochemical equations are kept without simplifications or approximations, and thus guarantees the validity of the model. Apart from the demonstration of the novel implementation, one additional purpose is to quantify its accuracy towards a 'full physical model', in our case in COMSOL Multiphysics, in terms of step time and number of meshes.

## 2. Model implementation

The discharge process in a lithium ion battery cell is described in Fig. 1, where lithium ions move from the negative electrode to the positive electrode inside the battery and electrons move from the negative electrode to the positive electrode through the outer circuit In the charge process, the opposite flows will occur. The basic steps shown in the figure include electrons moving in the solid phase, charge transfer and mass transport of lithium ions in both solid particles and in the liquid electrolyte. These processes can be modelled by a set of equations, which are listed in Table 1 [13]. In this work, these processes are represented with the electronic resistance in the solid $R_s$, charge transfer resistance $R_{ct}$, electrode potential U and electrolyte resistance $R_l$, which are connected in a transmission line structure. The transmission line structure was originally proposed by de Levie [24] assuming that the porous electrode consists of cylindrical particles. The transmission line structure can, however, be generally applied for a porous electrode if only the current distribution in the through-plane direction is considered, regardless of the particle shapes.

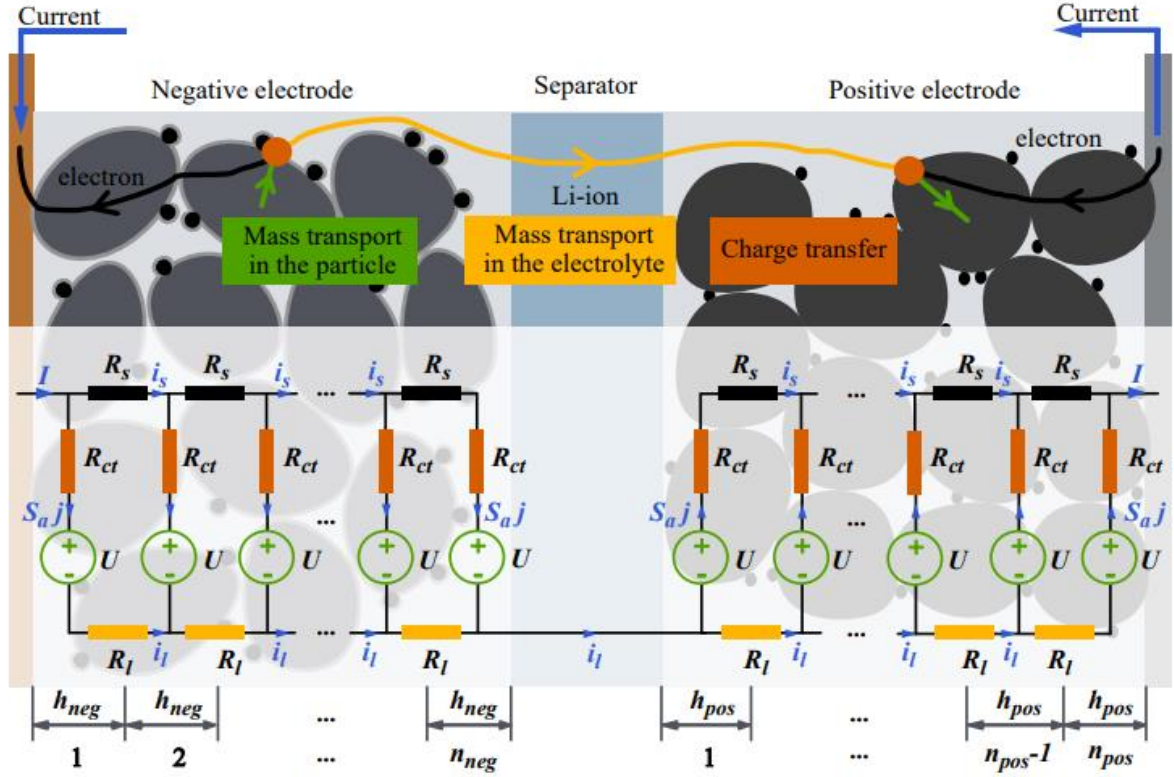

Figure 1: The discharge process in a lithium ion battery cell and the representation of the transmission line structure.

The electronic conduction follows Ohm's law in (2) and is modelled with a resistor $R_s$ in the circuit. The value of $R_s$ only depends on the electronic conductivity σs and the volume fraction of the solid matrix, so $R_s$ will remain the same during the simulation unless the model is coupled with, for example, a temperature or aging phenomenon. In commercial lithium-ion batteries, the electronic conductivity is improved by adding different types of carbon additives in the electrode, and $R_s$ can therefore in many cases be ignored. The charge transfer process is described by the Butler–Volmer equation in (4) and is represented by $R_{ct}$ in the circuit. Under a very small current, i.e. close to equilibrium, the Butler–Volmer equation can be linearized from a Taylor series

$$exp(x) = 1 + x + \frac{x^2}{2!} + \frac{x^3}{3!} + ... \approx 1 + x \tag{10}$$

and thus

$$j \approx j_0(1 + \frac{\alpha_a F \eta}{RT} - 1 + \frac{\alpha_c F \eta}{RT}) = j_0 \frac{F\eta}{RT} \tag{11}$$

where $j$ is the charge transfer current density per surface area, $j_0$ is the exchange current density, $F$ is the Faraday constant, $R$ is the gas constant, $T$ is temperature and η is the overpotential caused by the redox reaction. The linearization yields the charge transfer resistance for the active surface area $R'_{ct}$ as

$$R'_{ct} = \frac{\eta}{j} = \frac{RT}{j_0 F} \tag{12}$$

Table 1: Equations used in Newman's model.

| | Equations with boundary conditions | | Implementation |
|---|---|---|---|
| Potential 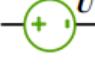 | $U = U(c_{s,surf})$ | (1) | Update the voltage source |
| 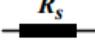 $R_s$ | $i_s = -\sigma_s^{eff}\nabla\Phi_s$ | (2) | Update the electrode resistance $R_s = -h\nabla\Phi_s/i_s$ |
| | $\nabla\Phi_s = -I/\sigma_s^{eff}$ at $x = 0$ and $x = L$ | | |
| 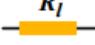 $R_l$ | $\vec{i}_l = -\kappa^{eff}\nabla\Phi_l + \frac{2\kappa^{eff}RT}{F}(1 + \frac{\partial ln f_A}{\partial ln c_l})(1 - t_+^0)\nabla ln c_l$ | (3) | Update the electrolyte resistance $R_l = -h\nabla\Phi_l/i_l$ |
| 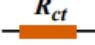 $R_{ct}$ | $j = j_0(exp\frac{\alpha_a F\eta}{RT} - exp\frac{-\alpha_c F\eta}{RT})$ | (4) | Update the charge transfer resistance $R_{ct} = \eta/S_a h j$ |
| | $j_0 = F k_c^{\alpha_a} k_a^{\alpha_c}(c_{s,max} - c_{s,surf})^{\alpha_a} c_{s,surf}^{\alpha_c} c_l^{\alpha_a}$ | | |
| Current distribution | $I = \vec{i}_s + \vec{i}_l$ | (5) | Mesh current method |
| | $S_a j = \nabla \cdot \vec{i}_l$ | (6) | Mesh current method |
| Concentration distribution | $\epsilon_l \frac{\partial c_l}{\partial t} = \nabla \cdot (\epsilon_l D_l^{eff}\nabla c_l) - \frac{\vec{i}_l \cdot \nabla t_+^0}{F} + \frac{S_a j(1 - t_+^0)}{F}$ | (7) | Finite difference method |
| | $\nabla c_l = 0$ at $x = 0$ and $x = L$ | | |
| | $\frac{\partial c_s}{\partial t} = D_s(\frac{\partial^2 c_s}{\partial r^2} + \frac{2}{r}\frac{\partial c_s}{\partial r})$ | (8) | Finite difference method |
| | $\frac{\partial c_s}{\partial r} = 0$ at $r = 0$, $\frac{\partial c_s}{\partial r} = -FD_s/j$ at $r = r_s$ | | |
| Porous electrode | $\sigma_s^{eff} = \sigma_s \epsilon_s^{1.5}$, $\kappa^{eff} = \kappa \epsilon_l^{1.5}$, $D_l^{eff} = D_l \epsilon_l^{1.5}$ | (9) | |

The charge transfer resistance for the electrode area $R_{ct}$ is scaled with the specific surface area $S_a$ and the length of the mesh element $h$

$$R_{ct} = \frac{R_{ct}'}{S_a h} \quad (13)$$

where $S_a$ can be estimated with the volume fraction of the solid $\epsilon_s$ and the particle radius rs

$$S_a = \frac{3\epsilon_s}{r_s} \quad (14)$$

The linearization above is only used to initialize $R_{ct}$ at the equilibrium state. During the simulation, $R_{ct}$ will be updated according to (4) with the corresponding local current density and lithium ion concentrations.

After the charge transfer process, a concentration gradient will be built up both in the particles and in the electrolyte. The concentration gradient together with the potential gradient are the driving forces for the mass transport process. The mass transport in the particle is expressed by Fick's law in (8) and the surface concentration $c_{s,surf}$ will determine the electrode potential $U$ in the circuit. Finally,

$R_l$ represents the resistance of the mass transport in the electrolyte, which is described with the concentrated electrolyte theory in (3) and (7). In the equilibrium state, there is no concentration gradient and $\nabla c_l = 0$ meaning that $R_l$ can be initialized with only the electrolyte conductivity $\kappa$. Later in the simulation, $R_l$ will be updated according to (3) based on the electrolyte current and concentration. All the transport parameter values are corrected with (9) thereby taking the porosity of the electrode into consideration

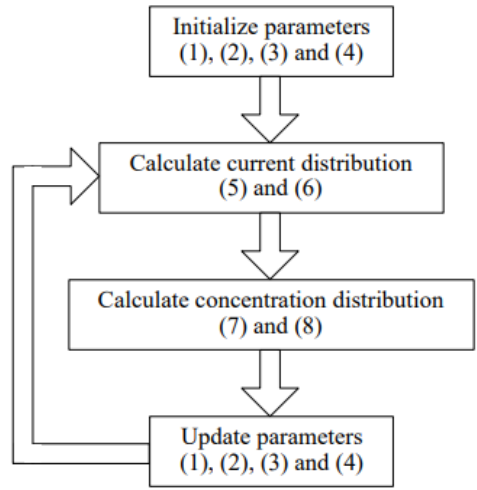

Figure 2: Calculation flow.

In this work, a decoupled quasi-dynamic simulation approach is implemented, where the current distribution is solved in an algebraic way and is considered to be static within the time step. With the formed static boundary conditions, the concentrations are solved dynamically, as shown in Fig. 2. The current distribution is solved with the mesh current method by the circuit structure, and the concentration distribution is solved with the finite difference method. The two methods are explained in the sections below.

**2.1. Solving the current distribution**

In a porous electrode, the current is distributed unevenly within the electrode, and the current density is normally higher close to the separator and lower close to the current collector. In this work, the current distribution is described with the transmission line circuit structure and solved with the mesh current method, instead of the (5) and (6). After initializing the component values in the circuit through (1)-(4), the resistance triangular matrix and voltage vector for the transmission line structure in Fig. 1 can be generated as

$$R_{matrix} = \begin{bmatrix} R_{s,1} + R_{ct,1} + R_{ct,2} + R_{l,1} & -R_{ct,2} & & & \\ -R_{ct,2} & R_{s,2} + R_{ct,2+3} + R_{l,2} & -R_{ct,3} & & \\ & & \ddots & \ddots & \ddots & \\ & & & -R_{ct,n} & R_{s,n} + R_{ct,n} + R_{ct,n+1} + R_{l,n} \end{bmatrix} \quad (15)$$

$$V_{vector} = \begin{bmatrix} U_1 + I(R_{ct,1} + R_{l,1}) - U_2 \\ U_2 + IR_{l,2} - U_3 \\ \ldots \\ U_n + IR_{l,n} - U_{n+1} \end{bmatrix} \quad (16)$$

where $I$ is the input current density. The current distribution can be calculated as

$$I_{vector} = R_{matrix}^{-1} V_{vector}, \quad (17)$$

where the elements in the current vector $I_{vector}$ refer to the mesh currents $I_1, I_2, ..., I_{1n-1}$ shown in Fig. 3. Thereby $i_s$, $j$ and $i_l$ can be calculated with Kirchhoff's circuit laws.

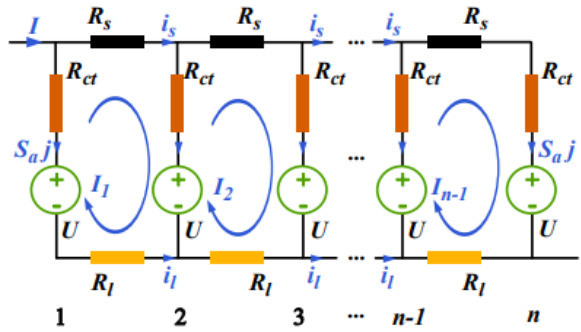

Figure 3: Mesh currents.

The resulting $i_l$ and $j$ will be used as input and boundary conditions in (7) and (8) to solve the concentration distribution, as shown in Fig. 4.

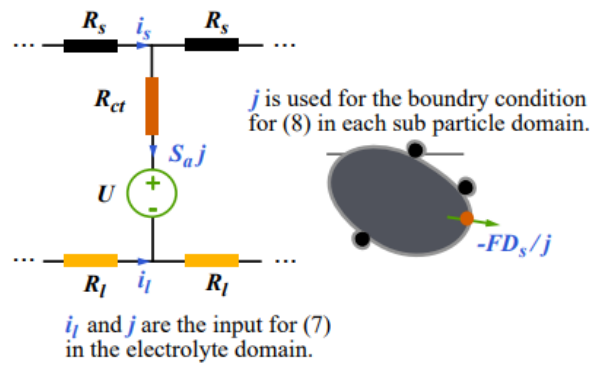

Figure 4: Solving the concentration distributions by using the result from the mesh current calculation.

## 2.2. Solving the concentration distribution

Within one time step, the current solved from the mesh current method is assumed to be constant, and the PDEs (7) and (8) can be solved numerically to obtain the concentration distribution. Both (7) and (8) are parabolic PDEs in one space dimension with Neumann boundary conditions. We have chosen the finite difference method for the spatial discretization because of the simple geometric feature for one dimensional problems. Other numerical methods, such as the finite element method, the finite volume method, and the difference potential method can however also be used [25].

There are two distinct difficulties in solving these two PDEs. In (7), the material porosity $\epsilon_l$ is discontinuous at the two material interfaces. For an accurate spatial discretization, the finite difference stencils should not cross the material discontinuities. In our method, we discretize (7) in each subdomain (negative electrode, separator and positive electrode) separately by finite difference operators with a summation-by-parts (SBP) property [26]. At the material interfaces, we impose physical interface conditions such that the concentration and its flux are continuous. These interface conditions as well as the Neumann boundary conditions are imposed numerically by the simultaneous-approximation-term (SAT) method [27]. As a result, the semi-discretized equations are energy stable. The resulting system of ordinary differential equations is solved by the MATLAB built-in function ode15s.

Equation (8) is reduced from the three dimensional heat equation in spherical coordinates when the solution is independent of the polar angle and azimuthal angle. The numerical difficulty here is the singularity at $r = 0$. In our method, we first multiply (8) by $r$ on both sides

$$r \frac{\partial c_s}{\partial t} = D_s \left( r \frac{\partial^2 c_s}{\partial r^2} + 2 \frac{\partial c_s}{\partial r} \right), \tag{18}$$

and then approximate the spatial derivatives in (18) by the SBP finite difference operators [28]. We again use the SAT method to impose the Neumann boundary conditions. In this case, the energy stable semi-discretized equations are a system of differential algebraic equations, and are also in this case solved by using the MATLAB built-in function *ode15s*.

After the current and concentration distributions are established, the resistance matrix will in the next step be updated according to (1)-(4) with the updated current and concentration values.

The MATLAB code is available at https://github.com/SiyangWangSE/CircuitModelLiionBattery.

## 3. Results and Discussion

### 3.1. Comparison to a physics-based model

With the implementation described above, an NMC-graphite lithium-ion battery — a widely used commercial battery chemistry especially for electric vehicles — is simulated at different discharge currents. The parameters used to describe the cell and its components are listed in Table 2 [29]. A simulation with the same cell parameters is performed with COMSOL Multiphysics as the comparison reference.

The oxidation reaction on the negative electrode side is

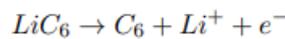

and on the positive electrode side the reduction reaction is

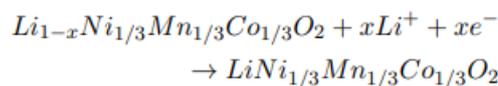

Table 2: Parameters used in the simulation.

| Parameters | Negative electrode | Separator (electrolyte) | Positive electrode |
|---|---|---|---|
| Length ($\mu$m) | 46.6 | 18.7 | 43 |
| $r_s$ ($\mu$m) | 6.3 | - | 2.13 |
| $\epsilon_l$ | 0.29 | 0.40 | 0.21 |
| $\epsilon_s$ | 0.49 | - | 0.57 |
| Bruggeman constant | 1.52 | 1.62 | 1.44 |
| Open circuit potential (V) | Look up table | - | Look up table |
| $\sigma_s$ (S/m) | 100 | - | 10 |
| $D_s$ (m$^2$/s) | $3 \times 10^{-14}$ | - | $5 \times 10^{-15}$ |
| $c_{s,max}$ (mol/dm$^3$) | 31.39 | - | 48.39 |
| $k_a = k_c$ | $2 \times 10^{-11}$ | - | $2 \times 10^{-11}$ |
| $\alpha_a = \alpha_c$ | 0.5 | - | 0.5 |
| $D_l$ (m$^2$/s) | | $2.5 \times 10^{-10}$ | |
| $\kappa$ (S/m) | | 1 | |
| $t_+^0$ | | 0.26 | |
| Mesh elements in the electrolyte domain | 20 | 20 | 20 |
| Mesh elements in the particle | 30 | - | 30 |
| Initial concentration (mol/dm$^3$) | $c_{s,max} \times 10\%$ | 1 | $c_{s,max} \times 90\%$ |

The simulated battery voltage profiles during discharge are presented in Fig. 5a, showing an excellent agreement with the results from COMSOL Multiphysics. The differences are shown in Fig. 5b in percentage. The difference is somewhat larger at higher currents, but still indicates a high accuracy for the transmission line-based model.

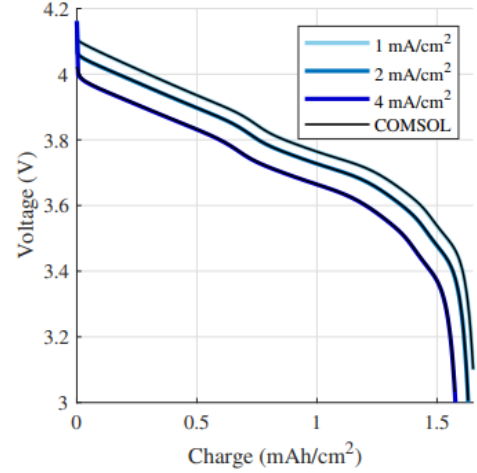

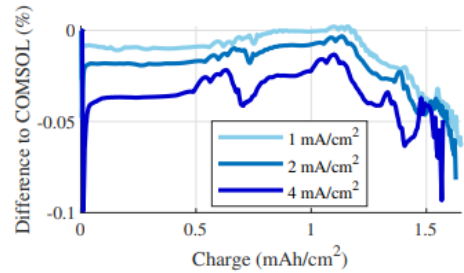

Figure 5: Simulated voltage profiles at different discharge current rates and a comparison with the result obtained from a simulation in COMSOL Multiphysics.

Besides the terminal voltage, the lithium ion concentration distributions are also critical in the simulation as the concentration is one key factor if aging phenomenon or mechanical stress are to be coupled in the model [30, 31]. The concentrations in the electrolyte is shown in Fig. 6a. At $t = 0$ s, the initial concentration in the electrolyte is 1 mol/dm$^3$. At $t > 0$ s, the battery starts to discharge and lithium ions travel from the negative electrode to the positive electrode and thus a concentration gradient starts to build up. This concentration gradient will quickly reach an equilibrium state in the beginning and then stay constant during the rest of the discharge time. Fig. 6b shows how the concentration at the particle surface changes with time. The initial lithium ion concentration is 10% in the negative electrode and 90% in the positive electrode which corresponds to a high SOC. During discharge, the negative electrode is being de-lithiated and the surface concentration decreases. As the current tends take the path of least resistance, the charge transfer current density is higher at the separator side (46.6 $\mu$m) than the current collector side (0 $\mu$m). Therefore the surface concentration decreases faster at the separator side. The opposite process happens in the positive electrode, i.e. the particles are lithiated during discharge and again with a higher current density at the separator side.

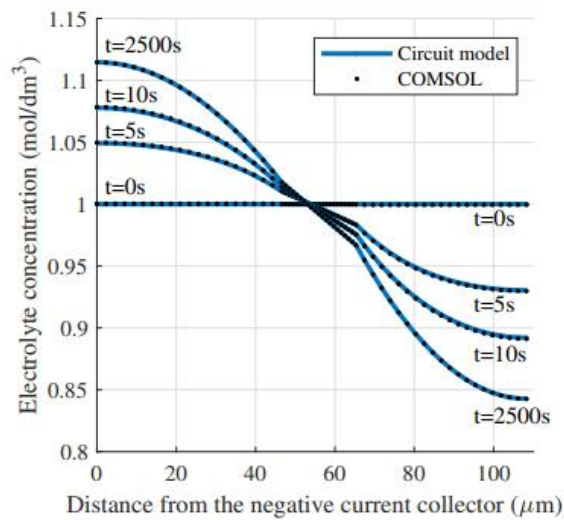

(a)

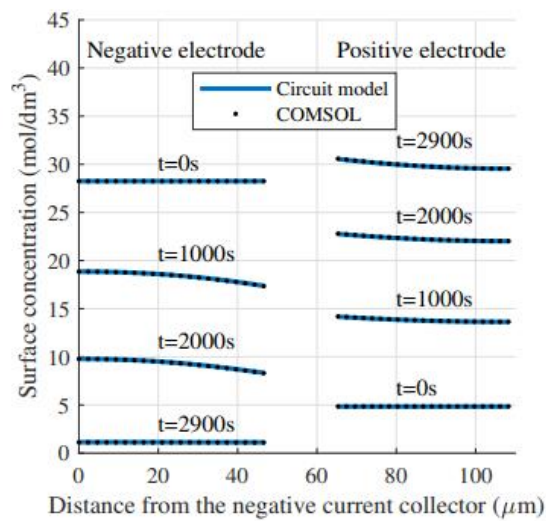

(b)

Figure 6: Electrolyte concentration and particle surface concentration changes with time under a discharge current density of 2 mA/cm$^2$

For the results shown in Fig. 5a, 20 mesh elements are used in the negative electrode, separator and positive electrode, respectively. In general, a finer mesh can give numerically more accurate results but on the other hand increases the computation time. In Fig. 7a the results with different meshing are presented using a discharge current of 2 mA/cm². For the case of 20 meshing elements, a comparison of 2nd order and 4th order space discretization is also shown. As can be seen, the 4th order space discretization provides a higher accuracy especially when the voltage profile starts to change rapidly at a low state of charge. However a minimum number of 8 mesh elements (9 grid points) is required in each domain to implement the 4th order space discretization.

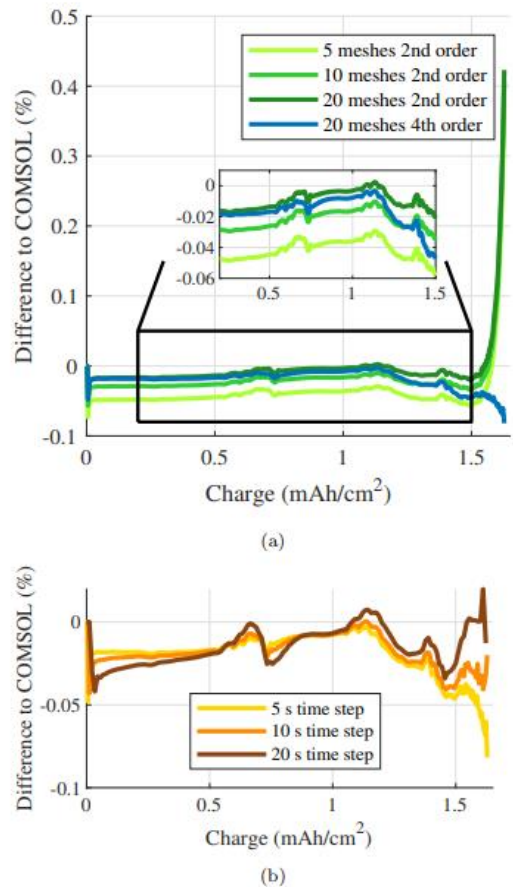

Figure 7: Performance of the model with different meshing sizes and different time steps under a discharge current density of 2 mA/cm².

Another factor that affects the accuracy and simulation time is the choice of the time step, especially when a quasi-dynamic approach is used, as in this work. For the results in Fig. 5a, the time step was 5 seconds and Fig. 7b shows the results for different choices of time steps. This indicates that under a constant current, the time step does not affect the result significantly.

### 3.2. Contributions to the overpotential

This current implementation with a transmission line structure serves as an interplay between physics-based models and equivalent circuit models. On one hand, it is built upon the first-principle electrochemical equations with physically based material parameters. On the other hand, it contains an illustrative yet simplistic structure that can be easily interpreted and straightforwardly solved. We exemplify here how to this transmission line circuit model can aid the understanding of the contributions of different processes to the overpotential.

In the previous calculations, the resistance elements $R_s$, $R_{ct}$ and $R_l$ represent the processes of electronic conduction, redox reaction and the mass transport in the electrolyte. The diffusion process in the solid has not been directly introduced, since the electrode potential is determined by the particle surface concentration, and which is reflected in the circuit. Here, however, we introduce the term $R_{diff,s}$ to represent the diffusion process in the solid, as shown in Fig. 8. With the presence of the resistance $R_{diff,s}$, the voltage source is determined from the average concentration in the active material particles instead of from the surface concentration.

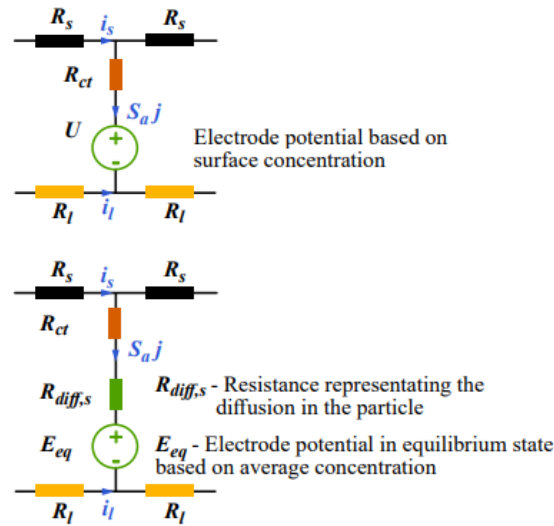

Figure 8: Equivalent resistance $R_{diff,s}$ to represent the overpotential caused by the diffusion in the solid.

A corresponding battery model is built using the parameters in Table 2 and discharged from 4.15 V with 2 mA/cm². When the battery is discharged to 3.6 V, the values of all the resistance elements are presented in Fig. 9 (note that the y-axis is in logarithmic scale). One observation is that the resistance caused by the mass transport in the electrolyte is a few orders of magnitude lower than the resistance caused by the diffusion in the solid, as the diffusion coefficient in the liquid $D_l$ is much higher than the diffusion coefficient in the solid $D_s$ ($D_l \gg D_s$). However the overpotential caused by mass transport in the electrolyte is significant, as demonstrated in Fig. 10.

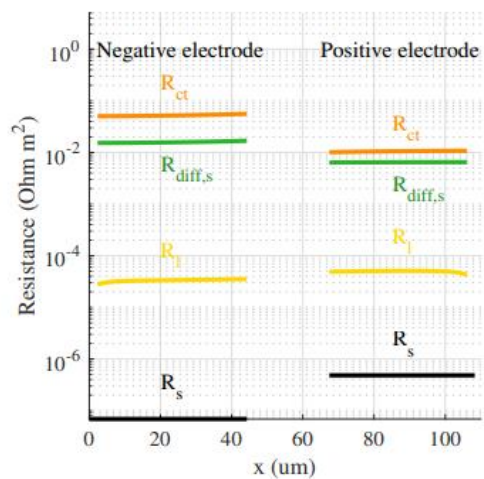

Figure 9: The resistance values in the transmission line structure when the cell voltage is discharged to 3.6 V under a discharge current density of 2 mA/cm². 20 mesh elements are used in each electrode. The y-axis is in logarithmic scale.

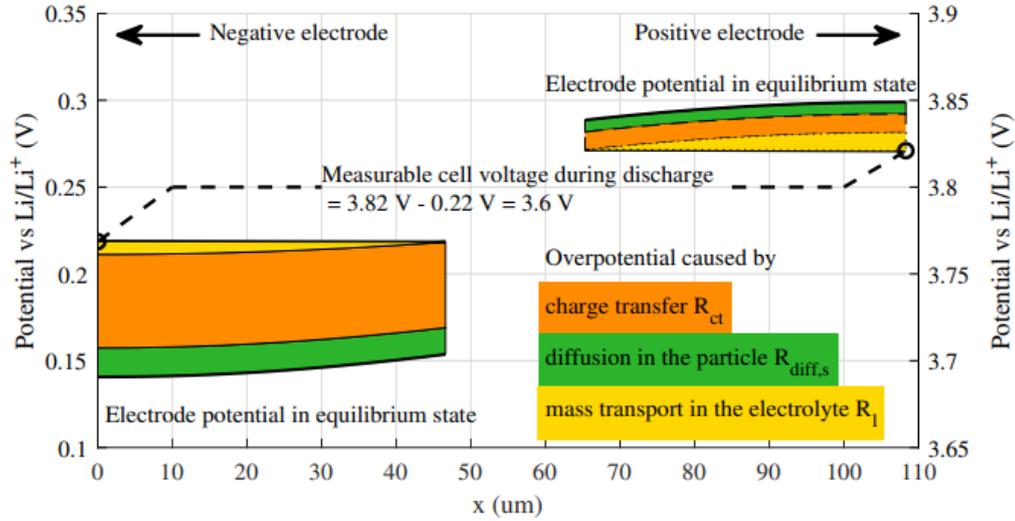

Figure 10: Overpotential caused by various processes in the electrode when the cell voltage is discharged to 3.6 V under a current density of 2 mA/cm².

One way to interpret this result is to compare the diffusion length. With 20 mesh elements in the positive electrode, the diffusion length is around 2 $\mu$m for both the liquid and solid phases, and the diffusion coefficient affects the resistance values ($R_l$ and $R_{diff,s}$). However, the total diffusion length for the electrolyte phase is the electrode thickness, which is around 20 times longer than the diffusion length in the solid. This, in turn, leads to the overpotential being significant. Similarly, another interpretation is to focus on the electrical circuit structure. The solid diffusion resistance elements $R_{diff,s}$ are connected in parallel, and only the branch current goes through each resistor. Contrarily, $R_l$ are connected in series, and the overpotential on each resistance element thereby ultimately adds up. As a consequence, the total overpotential becomes comparable to the overpotential caused by $R_{diff,s}$. This example shows that although the diffusion in the liquid is way faster than in the solid, the resulting overpotential caused by the electrolyte resistance cannot be ignored. Especially in the case of thick electrodes, the mass transport process in the electrolyte could be a limiting factor.

## 4. Conclusion

A new implementation of Newman's P2D model in MATLAB is presented in this work, constructed as an interplay between the physical and equivalent circuit models. A classic transmission line structure is used to replace the equations that describe the current distribution within the electrode. The concentration distributions are solved with the finite difference method and a decouple quasi-dynamic approach is used to combine the two solutions. The results from the circuit model agree very well with the result simulated with a commercial software COMSOL Multiphysics based on the finite element methods. This implementation closes the gap between physical and equivalent circuit models and is a useful tool to understand the processes inside the battery, as exemplified by the analysis of different contributions to the overpotential.

## Acknowledgment

The authors would like to thank Energimyndigheten for the financing of this work.

# Nomenclature

## Symbols

| | |
|---|---|
| $\alpha$ | Transfer coefficient |
| $\epsilon$ | Volume fraction |
| $\eta$ | Overpotential caused by the redox reaction, V |
| $\kappa$ | Electrolyte conductivity, S/m |
| $\Phi$ | Electrical potential, V |
| $\sigma$ | Electrode conductivity, S/m |
| $c$ | Lithium ion concentration, mol/m$^3$ |
| $c_{s,max}$ | Maximum concentration in the intercalation material, mol/m$^3$ |
| $c_{s,surf}$ | Lithium ion concentration on the particle surface, mol/m$^3$ |
| $D$ | Diffusion coefficient, m$^2$/s |
| $F$ | Faraday constant, 96485 s·A/mol |
| $f_A$ | Activity coefficient of the salt |
| $h$ | Element size in the meshing |
| $I$ | Given current density, A/m$^2$ |
| $i$ | Current density per electrode area, A/m$^2$ |
| $j$ | Charge transfer current density per surface area, A/m$^2$ |
| $j_0$ | Exchange current density, A/m$^2$ |
| $k_a$ | Anodic reaction rate constant |
| $k_c$ | Cathodic reaction rate constant |
| $L$ | Cell thickness, m |
| $n$ | Number of electrons involved in the redox reaction |
| $R$ | Gas constant, 8.314 J/mol·K |
| $r$ | Radial distance in the particle, m |
| $R_l$ | Electrolyte resistance, $\Omega$·m$^2$ |
| $R_s$ | Electronic resistance in the solid, $\Omega$·m$^2$ |
| $r_s$ | Radius of solid particles, m |
| $R_{ct}$ | Charge transfer resistance, $\Omega$·m$^2$ |
| $S_a$ | Specific surface area, m$^2$/m$^3$ |
| $T$ | Temperature, K |
| $t$ | Time, s |
| $t_+^0$ | Transference number of lithium ion |
| $U$ | Electrode potential, V |
| $V$ | Terminal voltage, V |

**Subscripts**

| | |
|---|---|
| *l* | Liquid phase |
| *neg* | Negative electrode |
| *pos* | Positive electrode |
| *s* | Solid phase |

**Superscripts**

| | |
|---|---|
| *eff* | Effective |